\documentclass[intlimits,twoside,a4paper]{article}

\usepackage[cp1251]{inputenc}

\usepackage{cmpj3}

\issue{2018}{21}{1}{13701}
\doinumber{10.5488/CMP.21.13701}

\title[$3d$-electrons contribution to cohesive energy]%
{$3d$-electrons contribution to cohesive energy of $3d$-metals}

\author{L.~Didukh}
\address{Ternopil Ivan Puluj National Technical University, Ternopil, Ukraine}

\date{Received July 18, 2017, in final form October 19, 2017}

\begin{document}

\maketitle

\begin{abstract}
In this paper a model for $3d$-subsystem of transition $3d$-metals has been proposed and used for calculation of the cohesive
energy dependent on $3d$-band filling of particular metal, its
bandwidth and effective intra-atomic interaction value. It has
been shown that the model enables one to explain the observed
peculiarities of cohesive energy effect on the atomic number. The nature of
two parabolic dependencies of cohesive energy on $3d$-band filling
has been clarified. The calculated values of cohesive energy are
close to those experimentally obtained for Sc-Ti-V-Cr-Mn-Fe series.
\pacs 71.10.Fd, 71.15.Nc
\end{abstract}

\section{Introduction}

\begin{figure}[!b]
\centering
\includegraphics[width=0.5\textwidth]{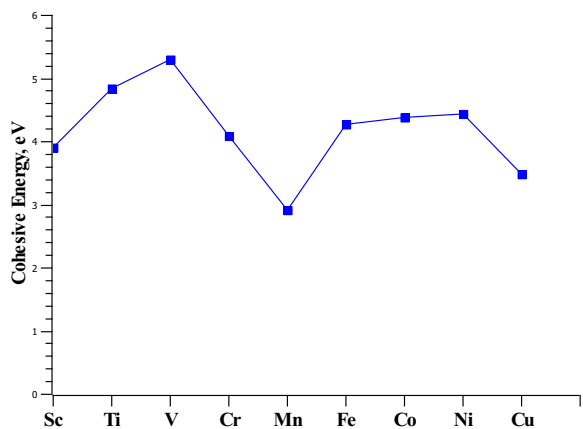} 
\caption{(Colour online) Dependence of
cohesion energy  on the atomic number for transition $3d$-metals.}
\label{coh_en}
\end{figure}

In figure~\ref{coh_en}, the experimental findings are presented for cohesive energy of
$3d$-metals depending on the atomic number (see~\cite{irkh07a}, figure~3.8).
Similar dependence on the atomic number is observed for melting
temperatures and boiling points of $3d$-metals~\cite{irkh07a,irkh07b}.
According to Friedel's theory, the cohesion energy of $d$-metals
is defined as a sum of energies for the occupied single-electron
states of a valence band~\cite{fri56}. If the simplest rectangular
density of states is used and equivalence of $d$-bands is assumed,
then a parabolic dependence of the cohesive energy on the number
of $d$-electrons follows. The parabolic dependence of cohesive
energy on the atomic number is in satisfactory agreement with the
experimentally found dependencies for transition $4d$- and
$5d$-metals. However, for $3d$-metals, as one can see from 
figure~\ref{coh_en}, there are substantial qualitative discrepancies between
the experimental and theoretical results. Clarification is necessary regarding this
``two-hump'' dependence and the anomaly associated with Mn.
It has been shown in papers~\cite{say77,ole81,acq83} that in non-degenerate 
(Hubbard) model, the cohesive energy
\begin{equation*}
 E_{\text{coh}} = \frac{n}{2}(2-n)w-\nu U,
\end{equation*}
where $w$ is the half-width of $s$-band, $U$ represents the
magnitude of intra-atomic Coulomb repulsion of two electrons with
opposite spins on the same lattice site, $n$ stands for the electron
concentration, $\nu$ is either the doubly occupied (with two
electrons on the same site) states concentration for $n<1$ or
empty sites (holes) concentration for $n>1$. As a consequence,
with interaction taken into account in the Hartree-Fock
approximation, two symmetrical parabolic dependencies $E_{\text{coh}}(n)$
(for $n<1$ and $n>1$) as well as the peculiarity for Mn are
explained.

An explanation of $E_{\text{coh}}(n)$ dependence asymmetry within the
$s$-band model was proposed in papers~\cite{did98, did00} by
taking into account the electron transfer caused by
electron-electron interaction, namely the correlated hopping of
electrons. 
This way we obtain two asymmetric parabolic dependencies for
$E_{\text{coh}}(n)$ with respect to $n=1$.
It is worth noting that the role of correlated hopping has been recently studied in papers~\cite{ashv14,gor14,ashv17}
within the context of electron interactions in strongly correlated electron systems.

 In the present work, developing the models \cite{irkh07a,irkh07b,fri56,say77,ole81,acq83,did98,did00,ashv14,gor14,ashv17,fri77}, 
the observed peculiarities of cohesive energy effect on the atomic number in $3d$-metals are interpreted
within the framework of the model with five-fold orbital
degeneracy of the band with taking into account the dependence of intraatomic interactions and
$3d$-band widths in $3d$-metals on their filling.

\section{The model}

1. We represent the Hamiltonian of the model generalized for
five-fold orbitally degeneracy (generalized Hubbard model) in the
following form
\begin{equation}
\label{Ham_gen}
H = {\sum\limits_{ijm\sigma}}'t_{im,jm}(n)a^{\dagger}_{im\sigma}a_{jm\sigma}+H_{\text{int}}.
\end{equation}
Here, the first term describes the delocalization energy of
$3d$-electrons, which provides the metallic bonding of atoms in
crystals, $t_{im,jm}$ is the delocalization integral of electrons,
$a^{\dagger}_{im\sigma}$, $a_{jm\sigma}$ are the operators of creation and
annihilation of electrons on the lattice site in state $m$ ($m=1,
2, 3, 4, 5$) with spin $\sigma$ ($\sigma = \uparrow, \downarrow$).
In the considered model, the transfer integral $t_{im,jm}(n)$
depends on the $3d$-electrons concentration $n_d$ in
transition $3d$-metal. This important peculiarity of the model
allows us to calculate the $3d$-electrons contribution in the
cohesive energy of particular $3d$-metals. This is the distinction
of the present model from the one proposed in paper~\cite{fri77}
where the energy band width was assumed independent of the band
filling and from ``non-Hubbard-type'' approach in papers \cite{tur08,phi96,xu15}.

$H_{\text{int}}$ describes the intra-atomic interactions in the
considered model and is a generalization of the Coulomb
interaction term
\begin{equation}
\label{U} U{\sum\limits_{i}}n_{i\uparrow}n_{i\downarrow}
\end{equation}
of the orbitally non-degenerate Hubbard model.

Let us assume that
\begin{equation}
\label{H_int} H_{\text{int}}=H_1+H_2+H_3\,,
\end{equation}
where
\begin{equation}
\label{H_1} H_1=U{\sum\limits_{im}}n_{im\uparrow}n_{im\downarrow}\,,
\end{equation}
\begin{equation}
\label{H_2} H_2 = U'{\sum\limits_{\substack{imm_1\sigma,\\ m\neq
m_1}}}n_{im\sigma}n_{im_1\bar{\sigma}}\,,
\end{equation}
\begin{equation}
\label{H_3} H_3 = U''{\sum\limits_{\substack{imm_1\sigma,\\ m\neq
m_1}}}n_{im\sigma}n_{im_1\sigma}.
\end{equation}
Here, $n_{im\sigma}$ is an operator of electron number in the state
$m$ with spin $\sigma$ on site $i$. The expression~(\ref{H_1})
describes the Coulomb repulsion of two electrons with opposite spins
in the same orbital $m$, the expression~(\ref{H_2}) represents the
Coulomb interaction of electrons in different orbital states
($\bar{\sigma}$ denotes the spin projection opposite to $\sigma$).
The expression~(\ref{H_3}) describes the Coulomb interaction of
electrons on different orbitals with taking into account the
intra-atomic exchange interaction, $U>U'>U''$  (for example,
see~\cite{lyo75}).

2. The energy spectrum of electrons, with $H_{\text{int}}$  taken into
account in the Hartree-Fock approximation in the absence of magnetic
ordering, is given by the expression
\begin{equation}
E_m({\bf k})=-\mu+t_m({\bf k}),
\end{equation}
where $t_m({\bf k})$ is a Fourier component of the transfer
integral $t_{im,jm}$. The chemical potential $\mu$ is renormalized
by taking into account the Coulomb and exchange interactions in
the Hartree-Fock approximation and is to be found from the
equation for $d$-electron concentration
\begin{equation}
n_d=\int_{-w}^{\mu} \rho(\epsilon)\rd \epsilon,
\end{equation}
where $\rho(\epsilon)$ is the electron density of states with
taking into account the orbital degeneracy.

3. In the orbitally non-degenerate model, the cohesive energy is
defined by the following expression
\begin{equation}
E_{\text{coh}} = -\frac{1}{N}{\sum\limits_{ij\sigma}}t(n)\langle
a^{\dagger}_{i\sigma}a_{j\sigma}\rangle -\nu U,
\end{equation}
where the first term is the electron delocalization energy and the
second one takes into account the delocalization energy lowering
by polar states. In the Hartree-Fock approximation, we have
\begin{eqnarray}
&&\nu = \frac{n^2}{4} \quad \textrm{for} \quad n>1,\label{lt1}\\
&&\nu = 1-n+\frac{n^2}{4} \quad \textrm{for} \quad
n<1\label{gt1}.
\end{eqnarray}
Using the model unperturbed rectangular density of electron states,
we obtain
\begin{equation}
\label{Ecoh} E_{\text{coh}} = \frac{1}{2w(n)}(w^2-\mu^2) -\nu U,
\end{equation}
where
\begin{equation}
\mu = w(n)(n-1).
\end{equation}
In the framework of orbitally non-degenerate model, the formula
(\ref{Ecoh}) reflects characteristic features of the experimentally
found dependency of cohesion energy on the atomic number in $3d$-metals
(\ref{coh_en}).

4. Let us extend the above considerations onto the model with
orbital five-fold degeneracy.
    We define the cohesive energy as
\begin{equation}
E_{\text{coh}} = {\int\limits_{-w}^\mu}\epsilon\rho(\epsilon)\rd \epsilon
-\nu U_{\text{eff}}(n_d),
\end{equation}
where $U_{\text{eff}}(n_d)$ is an effective intra-atomic interaction which
includes both Coulomb and intra-atomic (Hund's rule) exchange
terms, $\nu$ generalizes the expressions~(\ref{lt1}), (\ref{gt1}) of
the non-degenerate model.

Following Friedel, we take
\begin{equation}
\rho(\epsilon) = \frac{5}{w(n_d)}\,,
\end{equation}
where $w(n_d)$ is $3d$-band half-width which is dependent on the
band filling. Then,
\begin{equation}
\mu = \frac{w(n_d)n_d}{5}-w(n_d),
\end{equation}
thus,
\begin{equation}
\label{E_coh5} E_{\text{coh}} =
10w(n_d)\left[\frac{n_d}{10}-\left(\frac{n_d}{10}\right)^2\right]-\nu
U_{\text{eff}}(n_d).
\end{equation}
When correlation effects can be neglected, then the cohesive energy
maximum corresponds to the band center, as was obtained in the
Friedel's theory~\cite{fri56}.

\section{Application for $3d$-metals}
1. Let us interpret a general character of the $E_{\text{coh}}$
dependency by formula~(\ref{E_coh5}) on the energy parameters and the
band filling in $3d$-metals in terms of configurational (atom-band)
model of transition metal~\cite{did78a,did78b}. Accordingly, in the first
approximation by $n_d$ we mean the atomic values of the
corresponding element, then moving on to estimations for
transition metals, this statement will be corrected by taking into
account $s{-}d$-transitions, which cause deviations of $3d$-band
filling from atomic values of $n_d$.

Given the peculiarities of intra-atomic interactions, represented
by the expressions~(\ref{H_1})--(\ref{H_3}), for the case of $n_d<5$,
the Hund's polar states $3d^{n+1}$ will be relevant, for the case
of $n_d>5$, the non-Hund polar states (atomic configurations of
electrons with opposite spins on different orbitals) play the
role. The case of $n_d=5$ should be considered separately, because in
this case, the intra-atomic Coulomb interaction  is present at the same
orbital, which is considerably greater than interactions at
different orbitals. For these distinct regions of
$n_d$, the characteristic effective magnitudes of intra-atomic
interactions and averaged values of bandwidth will be selected.

Let us take for all $n_d<5$ the same values of $w(n_d)$ and
$U_{\text{eff}}(n_d)$ ($w_1$ and $U_1$, respectively). Besides,
generalizing the expression~(\ref{lt1}) for non-degenerate model,
we obtain
\begin{equation}
\nu_1U_{\text{eff}} =5\Big(\frac{n_d}{10}\Big)^2U_1.
\end{equation}
Here, the intra-atomic interaction is taken into account in the
Hartree-Fock approximation, $\nu_1$ is the polar $3d^{n+1}$-states
concentration. Therefore, for $n_d<5$
\begin{equation}
\label{E_lt5} \frac{E_{\text{coh}}}{10w_1} =
\left[\frac{n_d}{10}-\Big(\frac{n_d}{10}\Big)^2\right]-\Big(\frac{n_d}{10}\Big)^2
\frac{U_1}{2w_1}.
\end{equation}
In the case of $n_d>5$, we put values $w_2$ and $U_2$ of
formula~(\ref{E_coh5}) in correspondence with quantities $w(n_d)$
and $U_{\text{eff}}(n_d)$, and for formula~(\ref{gt1}), the corresponding
expression is
\begin{equation}
\nu_2U_{\text{eff}} =\left[5-n_d+5\Big(\frac{n_d}{10}\Big)^2\right]U_2.
\end{equation}
In accordance with electron-hole symmetry, here $\nu_2$ is the
``hole'' $3d^{n-1}$-states concentration. For the cohesive energy in
this case we obtain
\begin{equation}
\label{E_gt5} \frac{E_{\text{coh}}}{10w_2} =
\frac{n_d}{10}\left(1+\frac{U_2}{w_2}\right)-\left(\frac{n_d}{10}\right)^2\left(1+\frac{U_2}{10w_2}\right)-\frac{U_2}{2w_2}.
\end{equation}
From formulae~(\ref{E_lt5}) and~(\ref{E_gt5}) we have that
$E_{\text{coh}}(n_d)$ reaches its maximum values at
\begin{equation}
\label{lmin} n_d=\frac{10w_1}{2w_1+U_1}
\end{equation}
in the case of $n_d<5$ and for the case of $n_d>5$, the
corresponding value is
\begin{equation}
\label{rmin} n_d=\frac{10(w_2+U_2)}{2w_2+U_2}.
\end{equation}
One can see that taking into account the intra-atomic interaction
shifts the maximum of  $E_{\text{coh}}(n_d)$ to the left or to the right
from the band center.

The reasonable estimate of the intra-atomic interaction magnitude
is the halfbandwidth~\cite{geb97}. So, we take $U_1=w_1$ in
formula~(\ref{lmin}), which leads to $n_d\approx 3$. This
corresponds to the atomic value $n_d=3$ for vanadium. In analogous
way one can interpret the existence of the maximum value of
$E_{\text{coh}}$ by formula~(\ref{rmin}) for $n_d\approx 7$ (atomic value
of $n_d$ for cobalt). Hence, by formulae~(\ref{E_lt5})
and~(\ref{E_gt5}), two parabolic dependencies can be obtained with maxima specified
by formulae~(\ref{lmin}) and~(\ref{rmin}). This
can be put in correspondence with the experimentally found dependence of
cohesive energy on atomic number, shown in figure~\ref{coh_en}.

\begin{table}[!b]
\caption{The obtained results for cohesion energy. $E_1$ are cohesion
energies for atomic values $n_d$ and $W^{\text{exp}}$ used. For $E_2$
data for $n_d$ and $W^{\text{exp}}$ of table~\ref{tb1} were used, for
$E_3$ data for $n_d$ and $W$ of table~\ref{tb1} were used. $E^{\text{exp}}$
are experimental data~\cite{irkh07a,irkh07b}.} \label{tb2} \vspace{2ex}
\begin{center}
\begin{tabular}{|c|c|c|c|c|c|c|c|c|}
\hline\hline
 Metal    &$E_1$, eV  &$E_2$, eV &$E_2$, eV &$E_{\text{coh}}$, eV\strut\\
\hline\hline
 Sc       & 2.9       & 4.0      & 3.3      & 3.9         \strut\\
\hline
 Ti       & 4.6       & 5.4      & 5.0      & 4.85        \strut\\
\hline
 V        & 5.6       & 5.4      & 5.4      & 5.31        \strut\\
 \hline
 Cr       & 4.1       & 4.1      & 4.1      & 4.10        \strut\\
\hline
 Mn       &           &          & 2.6      & 2.92        \strut\\
\hline
 Fe       &           &          & 4.0      & 4.28        \strut\\
\hline\hline
\end{tabular}
\end{center}
\vspace{-3mm}
\end{table}
\begin{table}[!b]
\caption{Values for $3d$-band fillings given in~\cite{irkh07a,irkh07b},
$3d$-band widths $W$ given by~\cite{har80} and experimental
findings $W^{\text{exp}}$ by~\cite{kap91}.} \label{tb1} 
\vspace{2ex}
\begin{center}
\begin{tabular}{|c|c|c|c|c|c|c|c|c|}
\hline\hline
 Metal           &Sc    &Ti    &V      &Cr     &Mn    &Fe    &Co     & Ni \strut\\
\hline\hline
 $n_d$           & 1.76 & 2.90 &  3.98 & 4.96  & 5.98 & 6.94 & 7.86  &  8.97 \strut\\
\hline
 $W$ ($2w$), eV  & 5.13 & 6.08 &  6.77 & 6.56  & 5.60 & 4.82 & 4.35  &  3.78 \strut\\
\hline
  $W^{\text{exp}}$, eV   & 6.2  & 6.6  &  6.8  & 6.5   & 8.5  & 8.5  & 6.9   &  5.4 \strut\\
 \hline\hline
\end{tabular}
\end{center}
\end{table}

2. For an estimation of the cohesive energy magnitudes for
particular $3d$-metals, we  use the values for $3d$-band widths and
$3d$-band fillings given in~\cite{irkh07a} (table~2.1), where $W$ is the
bandwidth given by~\cite{har80} (table~20.4) and $W^{\text{exp}}$ are
experimental findings~\cite{kap91}.

One can see that there is a satisfactory agreement between
experimental values~\cite{har80} and calculated values~\cite{irkh07a,irkh07b} of the band widths for metals
of Sc-Ti-V-Cr series.

In table~\ref{tb2}, the results of cohesion energy calculation by
formula~(\ref{E_coh5}) with data from table~\ref{tb1} are
summarized. Here, $E_1$ are cohesion energies for the atomic values
$n_d$ ($s{-}d$-transitions neglected) and $W^{\text{exp}}$ used. $E_2$ is
cohesive energy from data for $n_d$ and $W^{\text{exp}}$ of
table~\ref{tb1}, $E_3$ are values of cohesive energy from data for
$n_d$ and $W$ of table~\ref{tb1}. $E^{\text{exp}}$ are the experimentally found
values for cohesive energy. In these calculations, an effective
intra-atomic interaction has been taken equal to half-width of
conduction band~\cite{geb97}.

\begin{figure}[!t]
 \centerline{\includegraphics[width=0.5\textwidth]{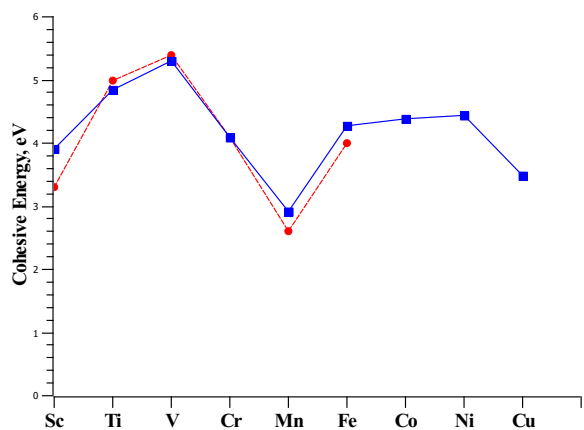}} 
 \caption{(Colour online) Values of cohesion
energy $E_3$ are marked red, experimental findings are marked
blue.} \label{coh_en3}
\end{figure}

Note that for Sc, Ti, V, Cr,  the values of $W$ and $W^{\text{exp}}$ are
close and reliable, while for Mn, Fe, Co, Ni, these values are
contradictory. For this reason, in table~\ref{tb1} the values for $E_1$
and $E_2$ for metals of the second group are omitted. The
calculated values are close to those experimentally obtained for
Sc-Ti-V-Cr-Mn-Fe sequence. Values $E_3$ for Co and Ni are
substantially lower than the experimental findings $E_{\text{coh}}$
(not listed in table~\ref{tb2}), this can signal of either the
inadequacy of the data listed in table~\ref{tb1} or the necessity
to go beyond the framework used for obtaining the
expressions~(\ref{E_lt5}) and~(\ref{E_gt5}), in particular, taking
into account the inter-orbital transitions of electrons and go
beyond the Hartree-Fock approximation. In figure~\ref{coh_en3},
the values of the cohesive energy are marked red. For $3d$-electrons in 
metals, positioned to the right of Mn, the description in terms of 
``configurational'' (atomic) model is more appropriate, as noted in 
chapter~3.5 of monograph~\cite{shch16}.
Summarizing, one can state that the proposed model not only
elucidates the nature of the two-hump dependence of cohesive
energy on the atomic number which is observed in transition $3d$-metals
and the peculiarity of $E_{\text{coh}}$ for Mn but also leads to the values
of cohesive energy close to those experimentally found for metals of
Sc-Ti-V-Cr-Mn-Fe series.

3. The results obtained from the calculation of $E_{\text{coh}}(n_d)$
allow one to interpret the dependencies of melting temperatures $T_{\text m}$
for $3d$-metals on the atomic number (analogous to the ones shown in
figure~\ref{coh_en}), as the melting temperature
\begin{equation}
T_{\text m} =\frac{0.04E_{\text{coh}}}{k_{\textrm{B}}}\,,
\end{equation}
where $k_{\textrm{B}}$ is Boltzmann constant~\cite{irkh07a,irkh07b}.

\section{Conclusions}
In this paper, a model for $3d$-subsystem of transition $3d$-metals
has been proposed and used for calculation of the
cohesive energy dependent on $3d$-band filling of a particular
metal, its bandwidth and the effective intra-atomic interaction
value.

It has been shown that the model allows one to explain the
observed peculiarities of cohesive energy on atomic number. The
nature of two asymmetric ``parabolic'' dependencies of cohesive energy on
$3d$-band filling has been clarified. The calculated values of
cohesive energy are close to the experimentally obtained for
Sc-Ti-V-Cr-Mn-Fe series. The obtained results can be extended for
explaining the peculiarities of melting temperatures of transition
$3d$-metals on their atomic number and other
systems~\cite{dov12,sko12,sko06,sko16} for which strong Coulomb
correlations determine the peculiarities of the energy spectrum.

\section*{Acknowledgements}
Fruitful discussions with Prof.~I.~Stasyuk are gratefully
acknowledged by the author.

\ukrainianpart

\title{Внесок $3d$-електронів у енергію зв'язку $3d$-металів}
\author{Л. Дідух}
\address{Тернопільський національний технічний університет імені Івана Пулюя, Тернопіль, Україна}

\makeukrtitle

\begin{abstract}
\tolerance=3000%
В роботі запропоновано модель підсистеми $3d$-електронів перехідних металів та застосовано її для розрахунку енергії зв'язку, залежної від заповнення $3d$-зони конкретного металу, ширини цієї зони та величини ефективної внутрішньоатомної взаємодії. Показано, що модель дозволяє пояснити спостережувані особливості залежності енергії зв'язку від атомного номера. Пояснено природу двох параболічних залежностей енергії зв'язку від заповнення $3d$-зони. Обчислені значення енергії зв'язку є близькими до отриманих експериментально для ряду Sc-Ti-V-Cr-Mn-Fe.
\keywords енергія зв'язку, $3d$-метали, міжелектронні взаємодії,
енергетичний спектр, орбітальне виродження
\end{abstract}

\end{document}